%% file: main.tex
\documentclass[manuscript]{acmart}
\AtBeginDocument{%
  \providecommand\BibTeX{{%
    \normalfont B\kern-0.5em{\scshape i\kern-0.25em b}\kern-0.8em\TeX}}}

\usepackage{booktabs} 

\usepackage[english]{babel}
\usepackage{moresize}
\usepackage{amsmath}
\usepackage{balance}
\usepackage{comment}
\usepackage{paralist}
\usepackage{bm}
\usepackage{pgfplots}
\pgfplotsset{compat=1.18} 
\usetikzlibrary{pgfplots.dateplot}
\usepackage{caption}
\usepackage{flushend}
\usepackage[english]{babel}
\usepackage[latin1]{inputenc}
\usepackage{mathrsfs}
\usepackage{graphicx}

\usepackage{amssymb}
\usepackage{amsfonts}
\usepackage{url}
\usepackage{longtable}
\usepackage{rotating}
\usepackage{multirow}
\usepackage{mathrsfs}
\usepackage{subfigure}
\usepackage{enumitem}
\usepackage[linesnumbered,algoruled,boxed,lined,ruled]{algorithm2e}
\usepackage{adjustbox}
\usepackage{hyperref}
\usepackage{pgfplots}
\usetikzlibrary{pgfplots.dateplot}
\definecolor{tblue}{RGB}{31,119,180}
\definecolor{torange}{RGB}{255,127,14}
\definecolor{tgreen}{RGB}{44,160,44}
\definecolor{tred}{RGB}{214,39,40}
\definecolor{tpurple}{RGB}{148,103,189}

\usepackage{array}
\usepackage{threeparttable}
\usepackage{tabularx}
\usepackage{booktabs}
\usepackage{makecell}
\newcolumntype{P}[1]{>{\centering\arraybackslash}p{#1}}
\newcolumntype{M}[1]{>{\centering\arraybackslash}m{#1}}

\newcommand{\hide}[1]{} 

\newcommand{\ie}{\textit{i}.\textit{e}.}
\newcommand{\eg}{\textit{e}.\textit{g}.}

\def\model{RCL}

\setcopyright{acmcopyright}

\copyrightyear{2023} 
\acmYear{2023} 
\setcopyright{acmlicensed}\acmConference[RecSys '23]{Seventeenth ACM Conference on Recommender Systems}{September 18--22, 2023}{Singapore}
\acmBooktitle{Seventeenth ACM Conference on Recommender Systems (RecSys '23), September 18--22, 2023, Singapore}
\acmPrice{15.00}
\acmDOI{10.1145/3604915.3608807}
\acmISBN{978-1-4503-9278-5/22/09}

\begin{document}

\title{Multi-Relational Contrastive Learning for Recommendation}


\author{Wei Wei}
\affiliation{%
  \institution{University of Hong Kong}
  \country{China}
}
\email{weiweics@connect.hku.hk}

\author{Lianghao Xia}
\affiliation{%
  \institution{University of Hong Kong}
  \country{China}
}
\email{aka\_xia@foxmail.com}

\author{Chao Huang}
\authornote{Chao Huang is the corresponding author.}
\affiliation{%
  \institution{University of Hong Kong}
  \city{}
  \country{China}}
\email{chaohuang75@gmail.com}






\begin{abstract}

Personalized recommender systems play a crucial role in capturing users' evolving preferences over time to provide accurate and effective recommendations on various online platforms. However, many recommendation models rely on a single type of behavior learning, which limits their ability to represent the complex relationships between users and items in real-life scenarios. In such situations, users interact with items in multiple ways, including clicking, tagging as favorite, reviewing, and purchasing. To address this issue, we propose the \underline{R}elation-aware \underline{C}ontrastive \underline{L}earning (\model) framework, which effectively models dynamic interaction heterogeneity. The \model\ model incorporates a multi-relational graph encoder that captures short-term preference heterogeneity while preserving the dedicated relation semantics for different types of user-item interactions. Moreover, we design a dynamic cross-relational memory network that enables the \model\ model to capture users' long-term multi-behavior preferences and the underlying evolving cross-type behavior dependencies over time. To obtain robust and informative user representations with both commonality and diversity across multi-behavior interactions, we introduce a multi-relational contrastive learning paradigm with heterogeneous short- and long-term interest modeling. Our extensive experimental studies on several real-world datasets demonstrate the superiority of the \model\ recommender system over various state-of-the-art baselines in terms of recommendation accuracy and effectiveness. We provide the implementation codes for the \model\ model at \url{https://github.com/HKUDS/RCL}.

\end{abstract}



\maketitle
\input{intro}
\input{model}

\input{solution}
\input{eval}
\input{relate}

\input{conclusion}
\input{appendix}

\clearpage
\bibliographystyle{ACM-Reference-Format}
\bibliography{sample-base}


\end{document}

%% file: intro.tex
\vspace{-0.05in}
\section{Introduction}
\label{sec:intro}
\vspace{-0.05in}


In personalized recommender systems, accurately capturing users' dynamic preferences over time is essential to provide effective recommendations. Recent advances in neural network architectures have led to the development of various models that can encode the evolving patterns of user-item interactions, leveraging deep learning techniques such as recurrent neural encoders~\cite{hidasi2015session}, convolution-based models~\cite{tang2018personalized}, and attention mechanisms~\cite{kang2018self}. More recently, recommender systems have employed the Transformer~\cite{sun2019bert4rec,liu2021augmenting} or Graph Neural Networks (GNNs)~\cite{wu2019session,ma2020memory,wang2020global} to achieve state-of-the-art recommendation performance. Although existing recommendation methods are effective, they typically rely on a single type of user-item interaction, such as clicks or purchases, which limits their ability to capture the dynamic multi-behavior interaction patterns that commonly occur in real-world recommender systems. 


In real-life recommendation scenarios, users frequently engage with items in diverse ways based on their evolving interests. These interactions can include various types of user behaviors, such as page views, adding items to favorites, and making purchases, which can reflect heterogeneous user intentions and relationships with items~\cite{jin2020multi,wei2022contrastive}. However, previous user embedding functions have typically relied on a single type of behavior modeling, which is insufficient for comprehensively capturing the diverse user intents and behavior heterogeneity. To address this limitation, researchers have proposed multi-behavior representations that can capture the various latent factors underlying user-item interactions and maintain distinct embedding spaces for different types of user behaviors in recommender systems~\cite{jin2020multi,xia2021knowledge,xia2021graph,chen2020efficient}. By leveraging multi-behavior representations, recommender systems can more effectively capture the complex relationships between users and items and provide more accurate recommendations.


Although it is essential to model behavior-aware user-item relationships in recommender systems, several key challenges remain to be addressed. Firstly, it is challenging to preserve the behavior-specific semantics that are pertinent to each type of user-item interaction in a dynamic manner. This is especially critical for multi-behavior recommendation, where it is necessary to distill heterogeneous item-level dependencies while jointly modeling short-term and long-term user interests. Secondly, learning informative and robust representations of multiplex user-item interactions requires a tailored modeling approach with a performant recommendation paradigm that can capture both the commonality and diversity of the different behaviors. While it is possible to embed behavior-specific semantics into individual latent vectors, it is also critical to understand the multi-behavior commonality underlying the global view of each user's dynamic preferences. By understanding this, it becomes possible to effectively model multi-behavior interactions and leverage the shared information across different types of user-item interactions.


\textbf{Contributions}. To overcome the challenges of modeling behavior-aware user-item relationships, we propose the \underline{R}elation-aware \underline{C}ontrastive \underline{L}earning (\model), which effectively captures heterogeneous and dynamic user intentions from multi-behavior data in recommendation. Our framework introduces a multi-relational graph encoder equipped with temporal context embedding to model behavior-aware short-term user interests and capture the dynamic and diverse nature of user-item interactions. Additionally, we propose a dynamic cross-relational memory network that incorporates heterogeneous item-item dependencies and learns user dynamic preferences with cross-behavior dependencies. Our dynamic multi-relational modeling approach enables us to characterize diverse user intents from a long-term perspective based on the interacted item sequence in dynamic environment. Furthermore, we design a multi-relational contrastive learning paradigm that enables \model\ to encode multi-behavior commonality and differentiate the behavior-aware preference of different users. Our framework learns informative and robust representations of user-item interactions across different types of behaviors and enhances the generalizability and robustness of our recommender.

To summarize, the key contributions of this work are presented as follows:

\begin{itemize}[leftmargin=10 pt, itemsep=1 pt, topsep=1 pt, partopsep= 1 pt]


\item Highlighting the importance of capturing the dynamic and heterogeneous nature of user-item relationships from multi-behavior data to make accurate recommendations.


\item Introducing a novel recommendation model called \model\ that combines dynamic cross-relational dependency modeling with a multi-relational contrastive learning paradigm to capture both short-term and long-term user interests. In addition, we perform a theoretical analysis of the \model\ model, which is presented in the Supplementary Section.


\item Conducting experiments on three real-world datasets and demonstrating the superior performance of the \model\ model. Performing ablation studies and in-depth model analysis to justify the rationale behind the model design.

\end{itemize}

%% file: model.tex
\vspace{-0.05in}
\section{Preliminaries}
\label{sec:model}
\vspace{-0.05in}

We consider a recommendation scenario with a set of users, $\mathcal{V}_u$, and a set of items, $\mathcal{V}_i$, where $u$ and $i$ represent the indices of users and items, respectively. We also introduce a set of behavior types, denoted by $\mathcal{B}$, with individual behavior index $b$. In this multi-behavior recommender system, we define \emph{target behaviors} as the types of user-item interactions that are of primary interest, such as purchase behavior in E-commerce platforms or like behavior in online video sites. Other types of interactions between users and items are considered \emph{auxiliary behaviors}, such as page view and add-to-favorite behaviors in E-commerce platforms or watch and review behaviors in online video sites. \\\vspace{-0.12in}


\noindent \textbf{Multi-Behavior Interactions}. We define a behavior-aware interaction sequence, denoted by $S_u$, for each user $u$ based on their historical interactions with items. The sequence is represented as $S_u = {(i_1, b_1), (i_2, b_2), ..., (i_{|S_u|}, b_{|S_u|})}$, where $b$ ($b\in \mathcal{B}$) represents the type of interaction between user $u$ and item $i$, and $|S_u|$ denotes the length of the multi-behavior sequence for user $u$. This representation captures the dynamic and heterogeneous nature of user-item relationships. \\\vspace{-0.12in}


\noindent \textbf{Problem Formulation}. The goal of the recommendation model is to predict the user's next preference based on their historical interactions with items and behavior types. To this end, the model takes as input the multi-behavior interaction sequence $S_u$ for each user $u$ with the awareness of interaction behavior types. The output of the model is the probabilities of user $u$ interacting with all candidate items in the next time step (after item $i_{|S_u|}$).


%% file: solution.tex
\vspace{-0.05in}
\section{Methodology}
\label{sec:solution}
\vspace{-0.05in}


This section describes the technical architecture of the \model\ model, which comprises three primary components. i) Short-term multi-behavior graph encoder, which utilizes multi-relational graph neural networks (GNNs) to capture the user's short-term interests. ii) Long-term user interest modeling module, which learns time-evolving multi-behavior preferences across different time slots. iii) Multi-behavior self-supervised learning module that enhances the user representation through behavior-aware contrastive augmentation.

\subsection{Short-Term Multi-Behavior Graph Encoder}
\label{sec:short}
\vspace{-0.05in}
To capture the dedicated behavior semantics for underlying user's short-term interests, we propose a multi-relational graph encoder to handle the heterogeneous item dependencies within each time slot.\\\vspace{-0.15in}

\vspace{-0.1in}
\subsubsection{ \bf Short-Term Multi-Behavior Graph}



We introduce the short-term multi-behavior graph $\mathcal{G}^b_t=(\mathcal{V}^b_t, \mathcal{E}^b_t, \textbf{M}^b_t)$, which represents the behavior-aware user-item interactions for behavior $b$ within the $t$-th time slot (e.g., day, week, or month). The graph consists of nodes $\mathcal{V}^b_t = \mathcal{V}{t,u}^b \cup \mathcal{V}{t,i}^b$, and edges $\mathcal{E}^b_t$ that represent the interactions between user nodes $\mathcal{V}^b_{t,u}$ and item nodes $\mathcal{V}^b_{t,i}$. Furthermore, we define the user-item interaction matrix $\textbf{M}^b_t$ based on the set of edges $\mathcal{E}^b_t$. Each entry $ \textbf{M}^b_{t,(u,i)}=1$ indicates that user $u$ has adopted item $i$ under behavior $b$ within the $t$-th time slot.

\begin{figure}
    \centering
    \includegraphics[width=1\columnwidth]{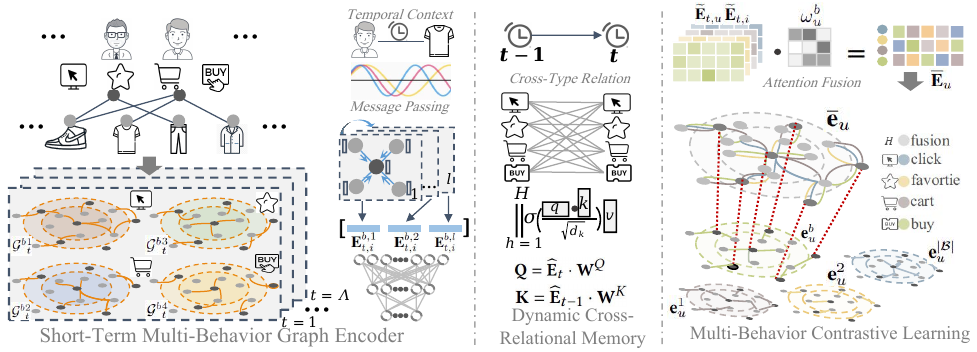}
    \vspace{-0.2in}
    \caption{The model flow of \model\ framework. The multi-behavior contrastive learning module augments the graph-enhanced dynamic memory network with the auxiliary behavior-aware self-supervision signals from both short-term and long-term user interests.}
    \vspace{-0.1in}
    \label{fig:framework}
\end{figure}

\subsubsection{ \bf Relation-aware Message Passing}
We propose a relation-aware message passing schema to model relational user data using lightweight graph neural networks, which is inspired by recent research efforts~\cite{he2020lightgcn}. This schema captures high-order dependencies among items through recursively propagating embeddings across graph layers. The relationa-aware message passing function from the ($l$-$1$)-th layer to the ($l$)-th layer is formalized as follows:
\begin{equation}
    \label{eq:item-gnn}
    \begin{split}
        \textbf{E}_{t, i}^{b,(l)} & = \sigma \left(\Gamma(\textbf{M}_{t}^b \cdot {\textbf{M}_{t}^b}^T) \cdot \textbf{E}_{t, i}^{b,(l-1)} \cdot \textbf{W}_{t,i}^{b,(l)} \right) \\
        \Gamma(\textbf{M}_{t}^b \cdot {\textbf{M}_{t}^b}^T) & = (\textbf{D}_{t}^b)^{-\frac{1}{2}} \cdot \textbf{M}_{t}^b \cdot ({\textbf{B}_{t}^b)}^{-1} \cdot {\textbf{M}_{t}^b}^T \cdot ({\textbf{D}_{t}^b)}^{-\frac{1}{2}}
    \end{split}
\end{equation}
\noindent where $\textbf{E}{t, i}^{b,(l)}$ represents the embedding of item $i$ at the $l$-th graph layer during the $t$-th time slot. The learnable transformation used during the message passing process is denoted as $\textbf{W}{t,i}^{b,(l)} \in \mathbb{R}^{d \times d}$. The authors use the PReLU activation function~\cite{he2015delving} $\sigma(\cdot)$ to introduce non-linearities during the message passing process. By stacking multiple embedding propagation layers, the authors preserve behavior-specific high-order item dependencies in the generated representations $\textbf{E}_{t, i}^{b,(L)}$, where $L$ denotes the number of propagation layers.

To address the issue of large value effects of embeddings during recursive propagation~\cite{wang2020next}, we incorporate the graph Laplacian normalized function of eigenvectors~\cite{kipf2016semi} $\Gamma(\cdot)$ into the message passing function. Specifically, we generate two diagonal degree matrices based on the interaction matrix $\textbf{M}t^b$. These matrices are $\textbf{D}{t}^{b} \in \mathbb{R}^{|\mathcal{V}_i| \times |\mathcal{V}u|}$ and $\textbf{B}{t}^{b} \in \mathbb{R}^{|\mathcal{V}_u| \times |\mathcal{V}_i|}$. The entries in these matrices are generated as follows:
\begin{equation}
    \begin{split}
        D_{t,(n,n)}^{b} = \sum_{n=1}^{|\mathcal{V}^{b}_{t, i}|}{\textbf{M}_{t,(n)}^{b}}; \qquad
        B_{t,(m,m)}^{b} = \sum_{m=1}^{|\mathcal{V}^{b}_{t, u}|}({{\textbf{M}_{t, (m)}^{b}})^T}
    \end{split}
\end{equation}
\noindent where $D_{t,(n,n)}^{b}$ and $B_{t,(m,m)}^{b}$ denote the elements of the $n$-th and $m$-th row of the diagonal matrices, respectively.
\\\vspace{-0.12in}

\noindent \textbf{Cross-Layer Aggregation.} To aggregate the embeddings encoded from each graph layer, we first concatenate all layer-specific embeddings and then feed the resulting representation into a feed-forward network. This network has a trainable projection matrix denoted as $\textbf{W}^b_{t,cat} \in \mathbb{R}^{(l \times d) \times d}$, which generates the final aggregated item embeddings $\textbf{E}_{t,i}^{b}$:
\begin{equation}
    \label{eq:item-agg}
    \begin{split}
        \textbf{E}_{t, i}^{b} = \left[ \textbf{E}_{t, i}^{b,1}:\textbf{E}_{t, i}^{b,2} :...:\textbf{E}_{t, i}^{b, l}  \right] \cdot \textbf{W}^b_{t,cat} ;\quad
        \textbf{E}_{t, i}^{b} = \frac{\textbf{E}_{t, i}^{b}}{||\textbf{E}_{t, i}^{b}||} 
    \end{split}
\end{equation}
\noindent By using the cross-layer aggregation method described above, the fused embeddings $\textbf{E}_{t, i}^{b}$ can preserve the behavior-aware short-term user preference at the $t$-th time slot under the behavior type of $b$.

\vspace{-0.1in}
\subsubsection{ \bf User Short-Term Interest Representation}
After encoding the item embeddings by exploring the high-order dependencies among items, we generate user representations that maintain the behavior-specific short-term interests. To do this, we aggregate embeddings propagated from the interacted items in the short-term multi-behavior graph $\mathcal{G}^b_t$. The embedding propagation from item $i$ to user $u$ is formally presented as follows:
\begin{equation}
    \begin{split}
        \textbf{m}_{u_t \gets i_t}^b = f(\textbf{e}_{i_t}^b, \textbf{e}_{u_t}^b) = \frac{1}{\sqrt{|\textbf{B}_{t,(u)}^{b}||\textbf{D}_{t,(i)}^{b}|}} \cdot  \textbf{W}_{t,u}^b \cdot \textbf{e}_{i_t}^b
    \end{split}
\end{equation}
\noindent $\textbf{e}{i_t}^b \in \mathbb{R}^{d \times 1}$ represents the item embedding within the $t$-th time slot under the behavior type of $b$, while $\textbf{W}{t,u}^b$ is the transformation matrix with dimensions of $\mathbb{R}^{d \times d}$. To generate the short-term interest representation of all users, we use a message passing scheme that can be presented in matrix form as follows:
\begin{equation}
    \label{eq:user-gnn}
    \begin{split}
        \textbf{E}_{t,u}^{b} = \sigma( \Gamma({\textbf{M}_{t}^b}^T) \cdot \textbf{E}_{t,i}^{b} \cdot \textbf{W}_{t,u}^{b}); \quad  \textbf{E}_{t, u}^{b} = \frac{\textbf{E}_{t, u}^{b}}{||\textbf{E}_{t, u}^{b}||};~~\Gamma({\textbf{M}_{t}^b}^T) = (\textbf{B}_{t}^b)^{-\frac{1}{2}} \cdot {\textbf{M}_{t}^b}^T \cdot (\textbf{D}_{t}^b)^{-\frac{1}{2}} 
    \end{split}
\end{equation}
\noindent To ensure the user representation $\textbf{E}_{t,u}^{b}$ corresponding to users' short-term interests at the $t$-th time slot under the behavior type of $b$ is properly normalized, we utilize the $\Gamma(\cdot)$ function.

\vspace{-0.1in}
\subsubsection{\bf Temporal Context Injection}
To capture the dynamic item dependencies in the short-term multi-behavior graph $\mathcal{G}^b_t$ and model the dynamic nature of user-item interactions, we propose to inject temporal contextual signals into the user representation. To achieve this, we build a temporal context encoder over the positional encoding technique in the Transformer architecture~\cite{vaswani2017attention}. Specifically, we use sinusoid functions to generate the relative time embedding for each user-item interaction edge. This enhances our short-term multi-behavior graph encoder with the capability of capturing behavior-aware interaction dynamics of users through temporal context injection.


\vspace{-0.1in}
\subsubsection{ \bf Priori-aware Embedding Initialization}
To address the potential issue of overfitting that may arise from relying solely on time slot-specific user-item interaction data in our short-term multi-behavior graph encoder, we develop a priori-aware embedding initialization strategy. This strategy injects the representations of the previous time slot $(t-1)$ into the embedding initialization of the current time slot $t$, with the aim of reducing training difficulty and deterioration~\cite{glorot2010understanding}. Formally, the priori-aware embedding initialization is defined as follows:
\begin{equation}
    \label{eq:gnn-init}
    \begin{split}
        \textbf{E}_{t,i}^{b, 0} = \left(\zeta* \textbf{E}_{t,i}^{b, 0} + (1-\zeta)* \textbf{E}_{t-1,i}^{b} \right) \cdot \textbf{W}_{\zeta}^{b} 
    \end{split}
\end{equation}
\noindent where $\textbf{E}_{t, i}^{b, 0}$ denotes the original input embeddings for the short-term multi-behavior graph $\mathcal{G}^b_t$ at the $t$-th time slot, while $ \textbf{E}_{t-1, i}^{b}$ represents the representations encoded from the previous $(t-1)$ time slot. The $\zeta$ and $\textbf{W}_{\zeta}^{b}$ variables are the learnable aggregation weight and projection matrix, respectively.

\vspace{-0.05in}
\subsection{Dynamic Cross-Relational Memory Network}
\label{sec:long}
\vspace{-0.05in}
The long-term multi-behavior dynamics in a multi-behavior sequential recommender system provide a holistic view of the diverse preferences of users over different time slots. To capture this global multi-behavior interest of users, we propose a dynamic cross-relational memory that learns the evolving cross-type behavior dependencies in a long-term manner.

\vspace{-0.05in}
\subsubsection{ \bf Multi-Behavior Embedding Tensor}
After encoding users' behavior-aware short-term interest with user ($\textbf{E}_{t, u}^{b}$) and item ($\textbf{E}_{t, i}^{b}$) representations, we construct a three-way tensor $\widehat{ \textbf{E}}_{t, u} \in \mathbb{R}^{|\mathcal{B}| \times |\mathcal{V}_u| \times d}$ and $\widehat{\textbf{E}}_{t, i} \in \mathbb{R}^{ 
 |\mathcal{B}| \times |\mathcal{V}_i| \times d }$ to represent the multi-behavior short-term interest. We stack the type-specific user and item embeddings as follows:
\begin{equation}
    \begin{split}
       \widehat{ \textbf{E}}_{t, u} = stack \left( \textbf{E}_{t, u}^{1},  \textbf{E}_{t, u}^{2},..., \textbf{E}_{t, u}^{|\mathcal{B}|}\right);~~\widehat{ \textbf{E}}_{t, i} = stack \left( \textbf{E}_{t, i}^{1},  \textbf{E}_{t, i}^{2},..., \textbf{E}_{t, i}^{|\mathcal{B}|}\right) \\
    \end{split}
\end{equation}
\noindent where $\textbf{E}_{t, u}^b, \textbf{E}_{t, i}^b$ denote the behavior-specific representations of users and items, respectively. As such, we can associate each time slot $t$ with a multi-behavior embedding tensor $\widehat{ \textbf{E}}_{t, u}$ and $\widehat{ \textbf{E}}_{t,i}$ for users and items, respectively.

\vspace{-0.1in}
\subsubsection{\bf Time-Evolving Cross-Type Behavior Dependencies}
In a dynamic multi-behavior scenario, different types of user behaviors are interdependent among time slots in a long-term perspective. For example, customers may view some products on online retail sites and make their purchase decisions one day later. To capture such evolving cross-type behavior dependencies across time slots, we develop a dynamic cross-relational memory network that explicitly learns the influence weights between time slot-specific behavior-aware representations.

Towards this end, we design a self-attention-based memory network to model the embedding correlations (\eg, $\textbf{E}_{(t-1), u}$--$\textbf{E}_{t, u}$) between adjacent time slots $(t-1)$ and $t$. In particular, we take the embedding $\textbf{E}_{t, u}$ as the input for query embeddings $\textbf{Q}_{t}$, and $\textbf{E}_{(t-1), u}$ as the inputs corresponding to the key and value embeddings $\textbf{K}_{t-1}$ and $\textbf{V}_{t-1}$. Our dynamic cross-relational memory encoder $\gamma(\cdot)$ is defined with the following scaled dot-product attention function over different time slot-specific behavior representations:
\begin{equation}
    \label{eq:self-attention-weights}
    \begin{split}
    \textbf{Z}_{t,t-1} = \gamma(\widehat{ \textbf{E}}_{t} \cdot \textbf{W}_t^Q, \widehat{ \textbf{E}}_{t-1} \cdot \textbf{W}_t^K, \widehat{ \textbf{E}}_{t-1});~~\textbf{Q}_{t} = \widehat{ \textbf{E}}_{t} \cdot \textbf{W}_t^Q; ~~~\textbf{K}_{t-1} = \widehat{ \textbf{E}}_{t-1} \cdot \textbf{W}_t^K
    \end{split}
\end{equation}
\noindent where $\textbf{Z}_{t,t-1}$ represents the aggregated multi-behavior embeddings of current time slot $t$ with the incorporation of cross-type behavior dependencies from the previous $(t-1)$-th time slot. $\textbf{W}_t^Q \in \mathbb{R}^{d\times d}$, $\textbf{W}_t^K \in \mathbb{R}^{d\times d}$ represent the linear transformation matrices corresponding to the query and key dimension, respectively. The learned cross-type behavior dependency matrix $\Phi \in  \mathbb{R}^{|\mathcal{B}| \times |\mathcal{B}|}$ which can be derived as $\Phi = \textbf{Q}_{t} \cdot \textbf{K}_{t-1}^T $. Each entry $\phi_{b,b'} \in \Phi$ indicates the estimated correlation between time slot- and behavior-specific representations $\textbf{E}_{t, u}^{b}$ and $\textbf{E}_{t-1, u}^{b'}$. Therefore, our dynamic cross-relational memory network $\gamma(\cdot)$ can be rewritten with the following form.
\begin{equation}
    \label{eq:qkv}
    \begin{split}
        \gamma( \textbf{Q}_t, \textbf{K}_{t-1}, \textbf{V}_{t-1}) = softmax \left( 
        \frac{ \textbf{Q}_{t} \cdot \textbf{K}_{t-1}^T }{ \sqrt{d/h}}  
        \right) \cdot \textbf{V}_{t-1}
    \end{split}
\end{equation}
The scale $\sqrt{d/h}$ is introduced to produce a softer attention distribution for avoiding gradient vanishing~\cite{vaswani2017attention}, and $h$ represents the number of head representations. With the design of our dynamic cross-relational memory network, our \model\ can preserve the dedicated time-evolving behavior dependencies across different types of user interactions. In our model implementation, we utilize the Broadcast Mechanism~\cite{van2011numpy} to improve the computational efficiency of tensor multiplication. With the learned evolving cross-type behavior dependencies, we refine the time slot-specific user and item embeddings as: $\widetilde{ \textbf{E}}_{t,u} = \mu(\widehat{ \textbf{E}}_{t, u} + 
        \textbf{Z}_{t,t-1, u})$ and $\widetilde{ \textbf{E}}_{t,i} = \mu (\widehat{ \textbf{E}}_{t, i} \oplus \textbf{Z}_{t,t-1,i}$), where $\mu(\cdot)$ indicates mean pooling operation.

\vspace{-0.05in}
\subsection{Multi-Relational Contrastive Learning}
\label{sec:beh-cl}
\vspace{-0.05in}
After encoding the heterogeneous user interest representations corresponding to the short-term ($\widehat{ \textbf{E}}_{t,u}, \widehat{ \textbf{E}}_{t,i}$) and long-term ($\widetilde{ \textbf{E}}_{t,u}, \widetilde{ \textbf{E}}_{t,i}$) perspectives, we propose to jointly capture the diverse multi-behavior dependencies and commonality in the multi-behavior sequential recommendation scenario.

\vspace{-0.1in}
\subsubsection{\bf Heterogeneous Behavior Aggregation}
We generate the multi-behavior representation ($\overline{\textbf{E}}_u$, $\overline{\textbf{E}}_i$) by aggregating type-specific behavior representations with a gating mechanism, which is formally presented as follows. Without loss of generality, $\overline{\textbf{E}}_u$ denotes the aggregated user representation for both short-term and long-term interests. Item embeddings $\overline{\textbf{E}}_i$ can be derived in an analogous way.
\begin{equation}
    \label{eq:beh-fusion}
    \begin{split}
        \omega_u^{b} =  \frac{ exp({ \textbf{E}_{u}^b  \textbf{ \textbf{W}}_f})}{ \sum_{b=1}^{|\mathcal{B}|}{ exp({ \textbf{E}_{u}^b \textbf{W}_f}})};~~~~~~~~
        \overline{ \textbf{E}}_{u} = \sum_{b=1}^{|\mathcal{B}|}\omega_{u}^b  \textbf{E}_{u}^b
    \end{split}
\end{equation}
\noindent where $\textbf{W}_f \in \mathbb{R}^{d}$ is the transformation matrix. $\omega_u^{b}$ represents the learned dependency weight of $b$-th type of user behaviors. By aggregating the type-specific behavior representations with the gating mechanism, the joint model can effectively preserve both the short-term and long-term multi-behavior preferences of users.

\vspace{-0.1in}
\subsubsection{\bf Cross-Relational Contrastive Self-Supervision}
In our \model\ model, we aim to simultaneously capture the commonality of individual users' multi-behavior patterns and the diversity of different users' multi-behavior preferences. To achieve this, we design the cross-behavior contrastive learning module, which enhances the multi-behavior dependency modeling with augmented self-supervision signals. Specifically, we generate contrasting views using type-specific behavior semantics and the fused multi-behavior pattern. Motivated by the property of contrastive learning, we incorporate cross-behavior contrastive self-supervision signals as auxiliary regularization in our recommendation framework, to jointly capture the commonality of individual users and the diversity of different users with respect to their multi-behavior preferences. Our \model\ performs behavior-level augmentation by pulling the type-specific behavior embedding ($\textbf{e}_{i}^{b} \in \mathbb{R}^{d \times 1}$) and multi-behavior representation ($\textbf{e}_{i} \in \mathbb{R}^{d \times 1}$) of the same user $u$ closer as positive pairs, and pushing the behavior embeddings of different users away as negative pairs. Formally, our cross-relational contrastive objective is defined as:
\begin{equation}
\label{eq:infoNCE}
    \begin{split}
        \mathcal{L}_{cl}^{b} = - \log  \frac{ exp({s( \textbf{e}_{u}^{b}, \overline{\textbf{e}}_u)/\tau})}{ exp({s( \textbf{e}_{u}^{b}, \overline{\textbf{e}}_u)/\tau}) + \sum\limits_{u\neq u'}{ \left( exp({s( \textbf{e}_{u}^{b}, \overline{\textbf{e}}_{u'})/\tau})  +  exp({s( \textbf{e}_{u}^{b}, \textbf{e}_{u'}^{b})/\tau})\right) }} 
    \end{split}
\end{equation}
\noindent where The temperature parameter $\tau$ is used to control the effect of mutual information estimation~\cite{chen2020simple}. We define the InfoNCE-based similarity function as $s(\textbf{e}_u^{b}, \overline{\textbf{e}}_u) = \textbf{e}_u^{b} \cdot \overline{\textbf{e}}_u/\Vert\textbf{e}_u^{b}\Vert\Vert\overline{\textbf{e}}_u\Vert$, which is measured by the dot product between $\ell_2$ normalized $\textbf{e}_u^{b}$ and $\overline{\textbf{e}}_u$.


\vspace{-0.05in}
\subsection{Model Inference Phase}
\vspace{-0.05in}
We define our optimized objective with the Bayesian Personalized Ranking (BPR) loss as follows:
\begin{align}
\label{eq:bpr-loss}
\mathcal{L}_{BPR} = \sum_{(u,i^+,i^-)\in O} - \log (\text{sigmoid} (\overline{e}_u \cdot \overline{e}_{i^+} - \overline{e}_u \cdot \overline{e}_{i^-} )) + \lambda || \Theta ||^2 
\end{align}
\noindent The pairwise training samples are $O ={ (u, i^+, i^-) | (u, i^+) \in \mathcal{\mathcal{E}}^+, (u, i^-) \in \mathcal{\mathcal{E}}^- }$, where $\mathcal{\mathcal{E}}^+$ and $\mathcal{\mathcal{E}}^-$ denote the corresponding observed and unobserved interactions of user $u$. The learnable hyperparameters are denoted by $\Theta$. To alleviate the overfitting, we apply $L_2$ regularization in our BPR loss. Our joint optimization objective is given below by integrating the BPR loss ($\mathcal{L}{BPR}$) with short-term ($\sum\limits{t}^{\Lambda}{ \sum\limits_{b}^{|\mathcal{B}|}{\mathcal{L}{cl}^{short}} }$) and long-term ($\sum\limits{b}^{|\mathcal{B}|}{\mathcal{L}_{cl}^{long}}$) contrastive objectives:
\begin{align}
\label{eq:total-loss}
\mathcal{L} = \sum\limits_{b}^{|\mathcal{B}|}{\mathcal{L}_{BPR}} + \alpha*\sum\limits_{b}^{|\mathcal{B}|}{\mathcal{L}_{cl}^{long}} + \beta*\sum\limits_{t}^{\Lambda}{ \sum\limits_{b}^{|\mathcal{B}|}{\mathcal{L}_{cl}^{short}} } 
\end{align}
\noindent where $\Lambda$ represents the number of time slots. $\alpha$ and $\beta$ are regularization strengths of contrastive objectives. The detailed training process of \model\ is elaborated in Algorithm~\ref{alg:learn}.

\begin{algorithm}[t]
\caption{The Learning Process of \model \ Framework}
\label{alg:learn}
\KwIn{
Behavior-aware interaction sequence $S_u = \{(i_1, b_1), (i_2, b_2),...,(i_{|S_u|}, b_{|S_u|})\}$. Short-term multi-behavior graph $\mathcal{G}^b_t=(\mathcal{V}^b_t, \mathcal{E}^b_t, \textbf{M}^b_t)$.
}
\KwOut{ 
Aggregated user/item representations $\overline{ \textbf{E} }_{i}$, $\overline{ \textbf{E} }_{u}$. 
The probability of the most likely next item $i_{|S_u|+1}$. \\
}
\textbf{Initialize:} 
Xavier initialized behavior-specific short term item embeddings $\textbf{E}_{t,i}^{b, 0}$. Parameters: i) Short-term multi-behavior graph encoder $\{ \textbf{W}_{t,i}^{b,(l)}, \textbf{W}_{t,u}^{b}, \textbf{W}^{b}_{\zeta}, \textbf{W}_{t,cat}^b \}$. ii)Dynamic cross-relational memory network $\{ \textbf{W}_t^Q, \textbf{W}_t^K \}$. iii) Behavior fusion transformation $\{ \textbf{W}_{f} \} $.\\
\For{ $epoch \leftarrow 0,1,...$}{
     Update learning rate scheduler. \newline
    \For{$step \leftarrow 0,1,...$}{
        \tcp{Short-Term Multi-Behavior Graph:}
        \For{$t \leftarrow 0,1,..., \Lambda$}{
            Get short term embeddings: $\textbf{E}_{t,u}^{b}, \textbf{E}_{t,i}^{b}$ $\longleftarrow$ $\mathcal{G}^b_t$, 
            \text{Eq.}1, \text{Eq.}3, \text{Eq.}5, \text{Eq.}6\\
            Prepare behavior aggregated embeddings for $\mathcal{L}_{cl}^{short}$: $\overline{\textbf{E}}_{t,u}$ $\longleftarrow$
            \text{Eq.}10\\
        }
        \tcp{Dynamic Cross-Relational Memory:}
        \For{$t \leftarrow 1,..., \Lambda$}{
            \text{Modeling time-evolving cross-type dependencies}: $\widetilde{\textbf{E}}_{t,u}^{b}, \widetilde{\textbf{E}}_{t,i}^{b}$  $\longleftarrow$ 
            \text{Eq.}7, \text{Eq.}8, \text{Eq.}9\\
            \text{Aggregate} $\widetilde{\textbf{E}}_{t,u}^{b}, \widetilde{\textbf{E}}_{t,i}^{b}$ \text{convey across $t$ and $b$ for $\mathcal{L}_{cl}^{long}$, $\mathcal{L}_{BPR}$ :} $\overline{\textbf{E}}_{u}, \overline{\textbf{E}}_{i}$ $\longleftarrow$
            \text{Eq.}10\\  
        }  
        \tcp{Cross-Behavior Contrastive Task \& Recommendation Task:}
        Get final multi-task objective $\mathcal{L}  \longleftarrow$ \
                $\mathcal{L}_{cl}^{b} \leftarrow$ 
        \text{Eq.}11 +        
        $\mathcal{L}_{BPR} \leftarrow$ 
        \text{Eq.}12 \\
        Gradient descent backpropagation. 
    }
}
\end{algorithm}  

\vspace{-0.05in}
\subsection{In-Depth Discussions on \model}
\vspace{-0.05in}
In this section, we present an in-depth analysis of our multi-behavior contrastive learning paradigm. Specifically, we discuss the benefits brought by the augmented behavior-aware self-supervised learning tasks from two dimensions: i) Cross-behavior mutual information maximization; ii) User interest representation discrimination. Additionally, we conduct a complexity analysis to study the efficiency of our \model\ framework.

\vspace{-0.1in}
\subsubsection{\bf Cross-Relational Mutual Information Maximization}
In our multi-relational contrastive learning framework, we propose to capture the commonality of individual users' multi-behavior patterns by maximizing the mutual information between the type-specific behavior embedding ($\textbf{e}_{i}^{b} \in \mathbb{R}^{d \times 1}$) and the multi-behavior representation ($\textbf{e}_{i} \in \mathbb{R}^{d \times 1}$). In particular, we define our information maximization function as $I( \textbf{z}_i, \textbf{z}_p)$, where $\textbf{z}_i, \textbf{z}_p$ represents the anchor and positive instance in the same hypersphere, respectively. Without loss of generality, the embedding $\textbf{z}$ is formally defined as: $\textbf{z} = \textbf{e}/\parallel \textbf{e} \parallel$. In our multi-relational contrasting scenario, the mutual information estimation is performed between the fused representation $\overline{ \textbf{E}}_{u}$ and behavior-specific embedding in $\{ \textbf{E}_{u}^{1},  \textbf{E}_{u}^{2},..., \textbf{E}_{u}^{b} \}$. These contrastive-based self-supervision signals are integrated into the BPR loss function to enhance the robustness of the user representation paradigm.




Motivated by the research in~\cite{van2018representation, hjelm2018learning, bachman2019learning, tian2020contrastive}, we formulate our augmented contrastive loss as a lower bound of the information maximization function $I(\cdot)$, which is defined as:
\begin{equation}
    I( \textbf{z}_i, \textbf{z}_p) \ge log(k) - \mathcal{L}_{cl}
\end{equation}
\noindent $k$ is the number of negative samples. The inequality suggests that a smaller value of $\mathcal{L}_{cl}$ results in a larger value of $I(\cdot)$. In other words, minimizing $\mathcal{L}{cl}$ is equivalent to maximizing the lower bound of mutual information in $I(\cdot)$.


The above situation can be extended to the multi-behavior modeling process. In addition to the equivalence between the contrastive objective and lower bound of information maximization, the self-supervised $\mathcal{L}_{cl}$ is closely related to the function $I(\cdot)$. This provides a theoretical basis for our multi-behavior recommender system, as follows:
\begin{equation}
    h^*_{\theta}( \textbf{z}_i, \textbf{z}_p) \propto \ \frac{p(\textbf{z}_i, \textbf{z}_p)}{p(\textbf{z}_i)p(\textbf{z}_p)} \propto \frac{p(\textbf{z}_i|\textbf{z}_p)}{p(\textbf{z}_i)}
\end{equation}
\noindent The optimal point $h^*$ of $h_{\theta}=\exp( s( \textbf{e}_{u}^{b}, \overline{\textbf{e}}_u)/\tau)$ is proportional to the density ratio between the joint distribution $p(\textbf{z}_i, \textbf{z}_p)$ and the product of marginals $p(\textbf{z}_i)p(\textbf{z}_p)$. This quantity is the point-wise mutual information, and the extended multi-behavior form can be implemented by optimizing the sum of a set of pairwise objectives \cite{tian2020contrastive}.

\vspace{-0.1in}
\subsubsection{\bf User Interest Representation Discrimination} 
Our developed contrastive paradigm not only captures the commonality among multiple behaviors but also enhances the diversity between individual nodes to mitigate over-smoothing \cite{liu2020towards,zhou2020towards}. 
During backpropagation, the gradient of the loss function $\mathcal{L}_{cl}$ with respect to the parameters of \model\ is computed and used to update the parameters, causing negative samples to move away from the anchor node $\overline{\textbf{e}}_u$ in the parameter space.
The following proportional relationship quantitatively illustrates how negative samples are pushed away from the anchor point as the similarity between samples increases:
    \begin{equation}
        \label{eq:prop-relat}
    \begin{split}
        c(x) \propto \sqrt{1-(x)^2} \cdot \exp{(x / \tau)}
    \end{split}
\end{equation} 
where node similarity score $x$ is computed using the normalized embeddings of the anchor node and negative sample. This score serves as the input to the function $c(\cdot)$. As the similarity $x$ increases, negative samples will have a greater gradient $c(x)$.
The temperature coefficient $\tau$ of the softmax function can adjust the gradient of hard negative samples. The similarity between each node and other nodes can be adjusted by the hyperparameter $\tau$.
As a result of the above analysis, we can conclude that multi-behavior contrastive learning can improve the discriminability of representations, thereby resolving the over-smoothing issue. A detailed discussion can be found in Appendix~\ref{sec:user_rep}.

\subsubsection{\bf Model Complexity Analysis}
The main time consumption in our \model\ framework
are from several key components: i) Short-term multi-relational graph encoder:
the computational cost of our graph neural architecture for item representation is $O(|\mathcal{B}|\times \Lambda\times L\times |\mathcal{E}^{b}_{t,M^TM}| \times d)$ for performing message passing across graph layers. Then for user, it is $O(|\mathcal{B}|\times \Lambda\times | \mathcal{E}^{b}_{t,M}| \times d)$. $|\mathcal{E}^{b}_{t, M^TM}|$, $|
\mathcal{E}^{b}_{t, M}|$ respectively represents the number of non-zero elements in the incidence matrix $\Gamma({\textbf{M}^b_t}^T)$ and $\Gamma({\textbf{M}^b_t}^T \textbf{M}^b_t)$, under the behavior type of $b$ during the time slot $t$. Here, $L$ denotes the number of graph propagation layers of item. 
The operations of concatenation and linear transformations for layer aggregation take $O(|\mathcal{B}|\times \Lambda\times L\times |\mathcal{V}^b_{i}| \times d)$. 
ii) Dynamic cross-relational memory network: 
The most computational cost for the cross-relational memory component comes from the self-attention operation with the time complexity of $O(\Lambda\times |\mathcal{V}^b| \times |\mathcal{B}|^{2}\times d)$ quadratic with the behavior number $|\mathcal{B}|$. iii) Multi-behavior contrastive learning: The cost of InfoNCE-based mutual information calculation is $O(d)$ and $O(batch \times d)$ for the numerator and denominator (in Eq.11
), respectively. Thus, our multi-relational contrastive learning paradigm takes $O(\Lambda\times |\mathcal{B}|\times | \mathcal{E}^{b} |\times d)$ per epoch. 


%% file: eval.tex
\section{Evaluation}
\label{sec:eval}

\begin{table}[t]
    \caption{Statistics of experimented datasets}
\vspace{-0.1in}
    \label{tab:data}
    \centering
    \begin{tabular}{cccccc}
    \hline
    Dataset & User \# & Item \# & Interaction \# & Sparsity & Interaction Behavior Types\\
    \hline
    Taobao & 31882 & 31232 & 167862 & 99.98\% & \{View, Favorite, Cart, Purchase\} \\ 
    IJCAI & 22438 & 35573 & 199654 & 99.98\% & \{View, Favorite, Cart, Purchase\}\\
    E-Commerce & 31021 & 1827 & 370386 & 99.35\% & \{Browse, Review, Purchase\} \\
    \hline
    \end{tabular}
\vspace{-0.1in}
\end{table}

\subsection{Experimental Setup}
\noindent \textbf{Datasets}. We perform model evaluations on three real-world datasets, and their statistics are shown in Table~\ref{tab:data}. i) \textbf{Taobao} collects four types of user behaviors from Taobao's recommender system. ii) \textbf{IJCAI} is an online retailing dataset released by IJCAI competition with four types of user online activities. iii) \textbf{E-Commerce} contains browse, review, and purchase behaviors of users in a real-life online retailer. Following the same settings in~\cite{mbgcn2020,xia2021graph}, our multi-behavior sequential recommender regards purchase as the target behavior and the other types of behaviors as auxiliary behaviors.

\noindent \textbf{Evaluation Protocols}. 
In our evaluation, we adopt two representative metrics, Hit Ratio (HR@N) and Normalized Discounted Cumulative Gain (NDCG@N) ($N=10$ by default), to measure the recommendation accuracy. Following existing sequential recommender systems~\cite{sun2019bert4rec}, we utilize the last interaction of each user as the positive sample for performance evaluation. Additionally, for each user, the last interacted item under the target behavior type is considered as the positive sample in the test set, and 99 non-interacted items are randomly sampled as negative instances.

\noindent \textbf{Baselines}. We compare our \model\ with various types of state-of-the-art recommendation methods.

\begin{itemize}[leftmargin=10 pt, itemsep=1 pt, topsep=1 pt, partopsep= 1 pt]
\item (i) CNN/Attention-based Sequential Recommendation Approaches: \textbf{Caser}~\cite{tang2018personalized} uses the convolutional filters to encode local item dependencies of user sequences. \textbf{SASRec}~\cite{kang2018self}, \textbf{TiSASRec}~\cite{li2020time}, \textbf{AttRec}~\cite{zhang2018next} are built on self-attention mechanism to capture correlations between temporally-ordered items. \textbf{Bert4Rec}~\cite{sun2019bert4rec} trains the bidirectional Transformer model with the cloze task for learning sequential patterns of user behaviors.

\item (ii) Hybrid Sequential Recommender Systems: \textbf{HGN}~\cite{ma2019hierarchical} is a hierarchical gating network that learns the item feature relevance for dynamic preference learning in recommendation. \textbf{Chorus}~\cite{wang2020make} considers both knowledge-aware item relations and temporal context in sequential recommendation.

\item (iii) GNN-based Sequential Recommendation Models: \textbf{SR-GNN}~\cite{wu2019session} introduces graph neural networks to model sequential behavior patterns over short item sequences. \textbf{MA-GNN}~\cite{ma2020memory} leverages GNNs to encode both short-term and long-term item dependencies. Furthermore, \textbf{HyperRec}~\cite{wang2020next} proposes to capture dynamic triadic item relationships using the hypergraph structures. \textbf{COTREC}~\cite{xia2021self} and \textbf{DHCN}~\cite{xia2021hyperself} are two state-of-the-art sequential recommender systems based on self-supervised learning paradigms.

\item (iv) Multi-Behavior Recommender Systems: \textbf{NMTR}~\cite{gao2019neural} and \textbf{DIPN}~\cite{guo2019buying} formalize the multi-behavior recommendation task with multi-task learning frameworks. \textbf{MBGCN}~\cite{jin2020multi}, \textbf{KHGT}~\cite{xia2021knowledge} and \textbf{MBGMN}~\cite{xia2021graph} design multi-behavior graph message passing schemes to model heterogeneous user-item interactions. \textbf{EHCF}~\cite{chen2020efficient} and \textbf{CML}~\cite{wei2022contrastive} generate additional supervision signals from auxiliary behaviors to boost the recommendation performance.

\end{itemize}

\begin{table}[t]
\centering
\setlength{\tabcolsep}{0.6mm}
\caption{Performance comparison of all methods on different datasets in terms of \emph{HR} \&  \emph{NDCG}.}
\vspace{-0.15in}
\small
\begin{tabular}{clccccccccccc}
\hline
    Data & Metric & \small{Caser}  & \small{AttRec} & \small{SASRec} & \small{TiSASRec} & \small{BERT4Rec} & \small{HGN}   & \small{Chorus} & \small{SR-GNN} & \small{MA-GNN} & \small{HyperRec} & \small{DHCN}    \\ \hline
\multirow{2}{*}{Tmall} & \small{H@10} & 0.321  & 0.328                                             & 0.319                                             & 0.322                                               & 0.329                                               & 0.283 & 0.335  & 0.318                                             & 0.331                                             & 0.333                                               & 0.321   \\
&\small{N@10} & 0.195  & 0.197                                             & 0.184                                             & 0.185                                               & 0.197                                               & 0.172 & 0.201  & 0.189                                             & 0.202                                             & 0.204                                               & 0.193   \\ \hline
\multirow{2}{*}{IJCAI} & \small{H@10} & 0.257  & 0.261                                             & 0.263                                             & 0.262                                               & 0.281                                               & 0.251 & 0.270  & 0.280                                             & 0.259                                             & 0.265                                               & 0.271   \\
& \small{N@10} & 0.146  & 0.147                                             & 0.148                                             & 0.148                                               & 0.155                                               & 0.136 & 0.149  & 0.151                                             & 0.149                                             & 0.151                                               & 0.148   \\ \hline
\multirow{2}{*}{E-Commerce} & \small{H@10} & 0.627  & 0.628                                             & 0.599                                             & 0.599                                               & 0.641                                               & 0.586 & 0.639  & 0.620                                             & 0.644                                             & 0.648                                               & 0.638   \\
& \small{N@10} & 0.381  & 0.387                                             & 0.368                                             & 0.368                                               & 0.392                                               & 0.359 & 0.402  & 0.373                                             & 0.411                                             & 0.413                                               & 0.391   \\ \hline \hline
   Data & Metric & \small{COTREC} & \small{NMTR}                                              & \small{DIPN}                                              & \small{MBGCN}                                               & \small{KHGT}                                                & \small{MBGMN} & \small{EHCF}   & \small{CML}                                               & \small{\textbf{RCL}}                                             & Imprv.                                              & p-value \\ \hline
\multirow{2}{*}{Tmall} & \small{H@10} & 0.330  & 0.362                                             & 0.325                                             & 0.381                                               & 0.391                                               & 0.419 & 0.433  & 0.543                                             & \textbf{0.597}                                             & 9.94\%                                              & $5.2e^{-5}$  \\
& \small{N@10} & 0.201  & 0.215                                             & 0.193                                             & 0.213                                               & 0.232                                               & 0.246 & 0.260  & 0.327                                             & \textbf{0.363}                                             & 11.01\%                                             & $4.9e^{-6}$  \\ \hline
\multirow{2}{*}{IJCAI} & \small{H@10} & 0.278  & 0.269                                             & 0.276                                             & 0.270                                               & 0.278                                               & 0.329 & 0.362  & 0.410                                             & \textbf{0.510}                                             & 24.39\%                                             & $9.6e^{-6}$  \\
& \small{N@10} & 0.153  & 0.156                                             & 0.151                                             & 0.138                                               & 0.145                                               & 0.176 & 0.207  & 0.235                                             & \textbf{0.312}                                             & 32.77\%                                             & $2.3e^{-5}$  \\ \hline
\multirow{2}{*}{E-Commerce} & \small{H@10} & 0.647  & 0.651                                             & 0.655                                             & 0.679                                               & 0.689                                               & 0.690 & 0.611  & 0.719                                             & \textbf{0.763}                                             & 6.12\%                                              & $8.7e^{-3}$  \\
& \small{N@10} & 0.405  & 0.408                                             & 0.397                                             & 0.414                                               & 0.434                                               & 0.432 & 0.413  & 0.427                                             & \textbf{0.470}                                             & 10.07\%                                              & $3.4e^{-4}$  \\ \hline
\end{tabular}
\begin{tablenotes}
\scriptsize
\item{*} To save space, we folded the table into two parts. Where the bolded columns are our results, followed by the \emph{improvment} and \emph{p-value} for the best results.
\end{tablenotes}
\label{tab:overall_results}
\vspace{-0.2in}
\end{table}

\noindent \textbf{Hyperparameter Settings and Implementation Details}. We implement our model in PyTorch and adopted the Xavier initializer~\cite{glorot2010understanding} for parameter initialization. The Cyclical Learning Rate strategy~\cite{smith2017cyclical} was used during the model training phase with the AdamW~\cite{loshchilov2017decoupled} optimizer. For fair comparison, the number of graph layers in all GNN-based models was selected from the range {1,2,3,4} to achieve the best performance. The temperature coefficient $\tau$ was tuned from {0.02, 0.035, 0.05, 0.07, 0.1, 0.3, 0.5, 0.7} in our multi-relational contrastive learning component. We further studied the influence of key hyperparameters in our model and reported the results in the Appendix.

\vspace{-0.05in}
\subsection{Recommendation Performance}
\vspace{-0.05in}
We report the performance comparison results in Table~\ref{tab:overall_results} and summarize the findings: (1) \model\ outperforms various baselines in all cases by achieving significant performance improvements. The ``imprv'' column indicates the relative performance improvement between \model\ and the best-performing baseline CML. Through the encoding of evolving dependencies across different types of behaviors, \model\ is able to simultaneously capture the dynamics of users' short-term and long-term interests by distilling the underlying heterogeneous interaction patterns. (2)  \model\ outperforms the compared sequential recommender systems by a large margin, which indicates that the incorporation of multi-behavior context is beneficial for disentangling the behavior heterogeneity of users. (3) Conducting dynamic contrastive learning with cross-behavior dependencies, \model\ learns better multi-behavior representations compared with state-of-the-art multi-behavior recommendation approaches by preserving both the behavior commonality and diversity of users. (4) The consistent performance improvements of our method on datasets with different sparsity degrees benefit from the incorporation of effective self-supervision signals generated by our contrastive learning paradigm.

\begin{table}[t]
    \small
    \caption{Ablation study on the effectiveness of components in \model.}
    \vspace{-0.1in}
    \centering
    \setlength{\tabcolsep}{1.4mm}
    \label{tab:ablation}
    \begin{tabular}{c|cc|cc|cc||c|cc|cc|cc}
    \hline
    Data & \multicolumn{2}{c|}{Tmall}      & \multicolumn{2}{c|}{IJCAI}      & \multicolumn{2}{c||}{E-Commerce} & Data & \multicolumn{2}{c|}{Tmall}      & \multicolumn{2}{c|}{IJCAI}      & \multicolumn{2}{c}{E-Commerce} \\
    \hline
    Metrics  & H@10 & N@10 & H@10 & N@10 & H@10 & N@10 & Metrics & H@10 & N@10 & H@10 & N@10 & H@10 & N@10 \\
    \hline
    \hline
    $w/o$-MBG  & 0.338 & 0.219 & 0.261 & 0.139 & 0.630 & 0.372 & $w/o$-CL  & 0.391 & 0.231 & 0.300 & 0.166 & 0.734 & 0.461\\
    \hline
    $r/w$-GRU & 0.361 & 0.210 & 0.276 & 0.148 & 0.675 & 0.417 & $w/o$-JBL & 0.473 & 0.291 & 0.367 & 0.205 & 0.747 & 0.469\\ 
    \hline
    \model & \textbf{0.597} & \textbf{0.363} & \textbf{0.510} & \textbf{0.312} & \textbf{0.763} & \textbf{0.470} & \model & \textbf{0.597} & \textbf{0.363} & \textbf{0.510} & \textbf{0.312} & \textbf{0.763} & \textbf{0.470} \\ \hline
    \end{tabular}
    \vspace{-0.1in}
\end{table}

\vspace{-0.05in}
\subsection{Ablation Study}
\vspace{-0.05in}
To validate the effectiveness of our proposed methodology, we conduct an ablation study from two perspectives: i) We examine whether the proposed technical modules contribute positively to the final performance. ii) we investigate whether the auxiliary user behavior data leads to performance improvements in our model.

\subsubsection{\bf Ablation Study on Proposed Modules}
We evaluate the efficacy of key components in \model\ with different variants: (1) \emph{w/o-JBL}: Instead of performing joint learning with multi-typed behavior supervision labels, we directly incorporate auxiliary behavior data as contextual features for user representation. (2) \emph{w/o-CL}: We further disable our multi-relational contrastive learning component to capture the behavior commonality and diversity. (3) \emph{r/w-GRU}: We replace our dynamic cross-relational memory network with gated recurrent unit (GRU) to encode behavior-specific embedding sequences. (4) \emph{w/o-MBG}: On the basis of \emph{w/o-CL}, we remove the behavior-aware graph neural encoder, to model the high-order connectivity over the multi-behavior user-item interaction graph within a specific time slot.

The evaluation results shown in Table~\ref{tab:ablation} demonstrate that our multi-relational learning with augmented short- and long-term contrastive objectives improves the generalization of recommender systems. Additionally, our cross-relational memory network is more effective than GRU in encoding long-term multi-behavior user interests. Furthermore, the multi-relational graph encoder has a positive effect on learning short-term heterogeneous user preferences compared with direct aggregation over ID-corresponding embeddings for collaborative modeling.

\subsubsection{\bf Ablation Study on Relation Multiplicity}
We compare the performance of our \model\ model when utilizing only a portion of the multiplex behavior data, as opposed to using all user behaviors. Results of this evaluation can be found in Table~\ref{tab:ab_data}. For each dataset, we remove each auxiliary behavior type such as views, favorites and cart behavior from the data before training and evaluating the performance of \model\ on the remaining dataset. We also remove all auxiliary behavior categories and only keep the target behavior type (Purchase). Our observations are as follows: First and foremost, auxiliary data is crucial. When utilizing only purchase data, the model performance is significantly worse than when auxiliary data is also included. This clearly highlights the benefits of including auxiliary user behavior data in recommendation models. Secondly, the importance of viewing and browsing data cannot be understated. As the auxiliary behavior with the largest amount, view and browse data are found to be the major contributors to performance improvements in our multi-behavior \model\ model. This underlines how auxiliary behavior data enhances the model through increased feature data and supervision signals. Finally, we found that favorite, cart, and review behaviors also have important contributions to model performance. Although the amount of these behavior records may be similar to purchase data, it is clear that they hold valuable user preference information.

\begin{table}[t]
\centering
\small
\caption{Performance of \model\ with different sub-sets of user behaviors.}
\label{tab:ab_data}
\vspace{-0.15in}
\setlength{\tabcolsep}{1.5mm}
\begin{tabular}{ccclcclcclcclcc}
\hline
    & H@10 & N@10 &  & H@10          & N@10         &  & H@10        & N@10       &  & H@10          & N@10  & & H@10 & N@10       \\ \hline
\multirow{2}{*}{Tmall}      & \multicolumn{2}{c}{$w/o$-View}   &  & \multicolumn{2}{c}{$w/o$-Favorite} &  & \multicolumn{2}{c}{$w/o$-Cart} &  & \multicolumn{2}{c}{Purchase} & & \multicolumn{2}{c}{\model}\\ 
\cline{2-3} \cline{5-6} \cline{8-9} \cline{11-12} \cline{14-15}
    & 0.463 & 0.264 &  & 0.547 & 0.327 &  & 0.534      & 0.319 &  & 0.370 & 0.230 & & 0.597 & 0.363\\ \hline
\multirow{2}{*}{IJCAI}  & \multicolumn{2}{c}{$w/o$-View}   &  & \multicolumn{2}{c}{$w/o$-Favorite} &  & \multicolumn{2}{c}{$w/o$-Cart} &  & \multicolumn{2}{c}{Purchase} & &\multicolumn{2}{c}{\model} \\ \cline{2-3} \cline{5-6} \cline{8-9} \cline{11-12} \cline{14-15}
                            & 0.355       & 0.197      &  & 0.417        & 0.234       &  & 0.463      & 0.269     &  & 0.305        & 0.177 & & 0.510 & 0.312  \\ \hline
\multirow{2}{*}{E-commerce} & \multicolumn{2}{c}{$w/o$-Browse} &  &  \multicolumn{2}{c}{$w/o$-Review}  &  & \multicolumn{2}{c}{-}  &  & \multicolumn{2}{c}{Purchase}  & & \multicolumn{2}{c}{\model}    \\ \cline{2-3} \cline{5-6} \cline{8-9} \cline{11-12} \cline{14-15}
                            &   0.711        & 0.441  &  & 0.732       & 0.446       &  & - & - & & 0.677      & 0.411  & & 0.763 & 0.470 \\ \hline
\end{tabular}
\vspace{-0.2in}
\end{table}

\vspace{-0.05in}
\subsection{  Alleviating the Data Sparsity Issue}
\vspace{-0.05in}

Our above theoretical discussion analyzes the benefits of our multi-relational contrastive learning paradigm in capturing the multi-behavior commonality and diversity. To be specific, the mutual information maximization between positive samples is helpful to preserve the common characteristics among different types of behaviors. Additionally, contrastive learning with negative samples can push the hard negative samples away to get distinguishable user embedding, so as to encode the behavior diversity of different users and alleviate the over-smoothing problem of our graph neural model.


Therefore, contrastive learning can improve the quality of representation and alleviate the problem caused by the scarcity of data. To this end, we select users whose interaction number on the IJCAI dataset in \{<5, <15, <35, <60\} for training and testing. In order to eliminate the influence of training data and simulate a real sparse data scenario, we carry out the evaluation of our model under the setting of $w/o$-JBL, and compare it with two best-performed baselines (HyperRec and KHGT) from the lines of sequence-based models and multi-behavior recommender systems, respectively.


As shown in Table~\ref{table:data_sparsity}, it can be observed that under different sparsity degrees of user interaction data, $w/o$-JBL with contrastive learning task will get better results than $w/o$-CL. Moreover, the performance gap between our contrastive learning method and other baselines becomes larger with the higher sparsity degrees of interaction data, which again justifies the effectiveness of our \model\ in addressing the data scarcity for recommender system.

\begin{table}[h]
\centering
\small
\vspace{-0.1in}
\caption{Performance with different number of interactions in terms of HR@5@20 \&
NDCG@5@20.}
\vspace{-0.1in}
\setlength{\tabcolsep}{0.45 mm}
\begin{tabular}{c|cccccccccccccc||ccccccccccccc}
\hline
                            \multirow{2}{*}{Metric} &   \multirow{2}{*}{Model}   &  & \multicolumn{2}{c}{\textless{}5} &  & \multicolumn{2}{c}{\textless{}15} &  & \multicolumn{2}{c}{\textless{}35} &  & \multicolumn{2}{c}{\textless{}60} & & \multirow{2}{*}{Model} &  & \multicolumn{2}{c}{\textless{}5} &  & \multicolumn{2}{c}{\textless{}15} &  & \multicolumn{2}{c}{\textless{}35} &  & \multicolumn{2}{c}{\textless{}60} \\ 
                             \cline{4-5} \cline{7-8} \cline{10-11} \cline{13-14} \cline{18-19} \cline{21-22} \cline{24-25} \cline{27-28}
                             &      &  & @5              & @20            &  & @5              & @20             &  & @5              & @20             &  & @5              & @20 &&& & @5              & @20            &  & @5              & @20            &  & @5              & @20            &  & @5              & @20                        \\ \hline \hline
HR & \multirow{2}{*}{HyperRec}    &  & 0.093                & 0.305               &  & 0.131          & 0.347          &  & 0.162          & 0.402          &  & 0.318          & 0.591     & &\multirow{2}{*}{KHGT}    &  & 0.114          & 0.337         &  & 0.176          & 0.417          &  & 0.220          & 0.470          &  & 0.409          & 0.636     \\
                             NDCG &  &  & 0.073                & 0.111               &  & 0.080          & 0.140          &  & 0.093          & 0.166          &  & 0.207          & 0.287    &  &&  & 0.075          & 0.137         &  & 0.116          & 0.172          &  & 0.146          & 0.210          &  & 0.237          & 0.317      \\ \hline
HR & \multirow{2}{*}{$w/o$-cl}      &  & 0.133          & 0.367         &  & 0.186          & 0.415          &  & 0.227          & 0.492          &  & 0.409          & 0.773    & &\multirow{2}{*}{$w/o$-JBL}    &  & 0.181          & 0.427         &  & 0.217          & 0.458          &  & 0.255          & 0.556          &  & 0.364          & 0.773     \\
                             NDCG & &  & 0.082          & 0.148         &  & 0.124          & 0.188          &  & 0.151          & 0.226          &  & 0.291          & 0.390 &  &  & & 0.117          & 0.187         &  & 0.146          & 0.214          &  & 0.179          & 0.2621         &  & 0.201          & 0.315         \\ \hline
\end{tabular}
\label{table:data_sparsity}
\vspace{-0.15in}
\end{table}

\vspace{-0.05in}
\subsection{Hyperparameter Sensitivity Analysis}
\label{sec:hyperparam}
\vspace{-0.05in}

\begin{figure}[t]
    \centering
    \includegraphics[width=0.3\columnwidth]{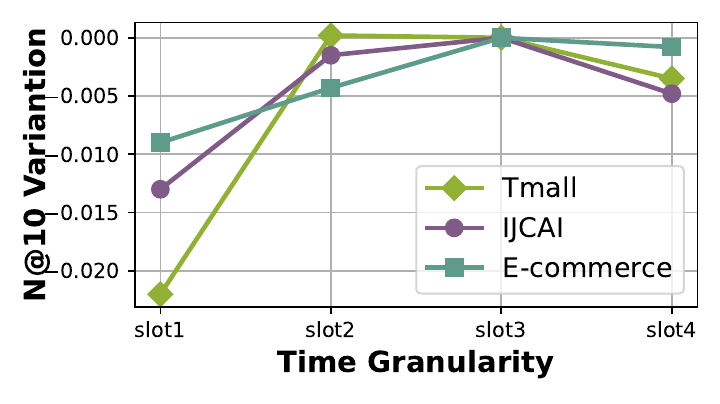}\,\,\,\,
    \includegraphics[width=0.3\columnwidth]{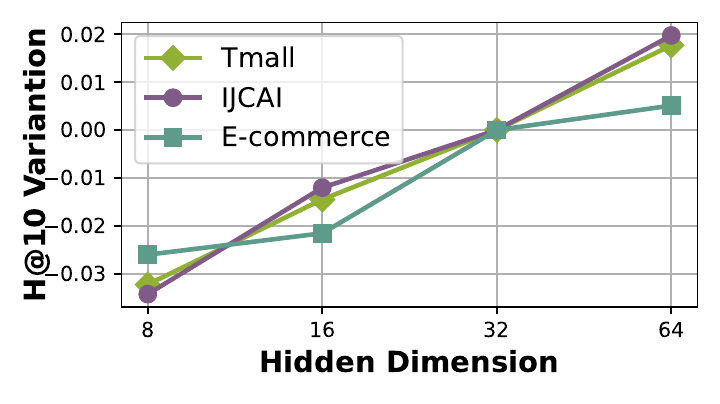}\,\,\,\,
    \includegraphics[width=0.3\columnwidth]{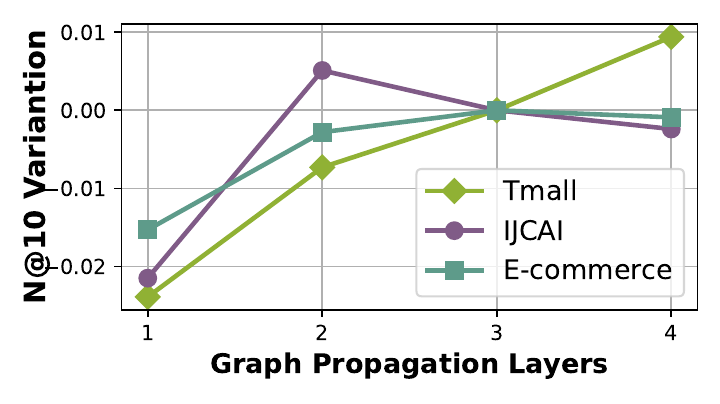}
    \vspace{-0.15in}
    \caption{Hyperparameter analysis of \model}
    \vspace{-0.2in}
    \label{fig:hyper-para}
\end{figure}

\noindent \textbf{Short-Term Time Granularity.}
To construct the short-term multi-behavior graphs, we tune the parameter of the time granularity from different time ranges due to the time span of different experimental datasets. In particular, we select the time granularity of Tmall, IJCAI, and E-commerce dataset from \{1, 3, 5, 9\} days, \{1, 3, 6, 12\} months, and {1, 2, 4, 12} weeks, respectively. From the evaluation results shown in Figure~\ref{fig:hyper-para}, we observe that shorter time periods (i.e., higher time granularity) can lead to overfitting in modeling short-term behavior-aware user preferences.

\noindent \textbf{Hidden Representation Dimensionality.} 
We conduct a parameter sensitivity analysis to investigate the impact of hidden dimensionality on our model's representation performance. Specifically, we vary the embedding size from a range of \{8, 16, 32, 64\}. Our results demonstrate that the performance on Tmall and IJCAI datasets improves significantly with an increase in the embedding size. This is due to the stronger representation power of larger embeddings.

\noindent \textbf{\# of Graph Propagation Layers.}
In order to capture short-term multi-behavior user interests, we have designed a behavior-aware message-passing scheme that refines user/item embeddings by injecting multi-typed behavior context. To explore the effect of model depth, we have selected the number of GNN layers from a range of \{1, 2, 3, 4\}. Our observations indicate that deeper graph models may be beneficial for modeling the high-order collaborative effects on Tmall and E-commerce datasets. However, we also found that stacking more embedding propagation layers can introduce noise in representation refinement on the E-commerce dataset.

\vspace{-0.05in}
\subsection{In-Depth Investigation of \model\ Model}
\label{sec:in_depth_exp}
\begin{figure}
    \centering
    \includegraphics[width=0.94\columnwidth]{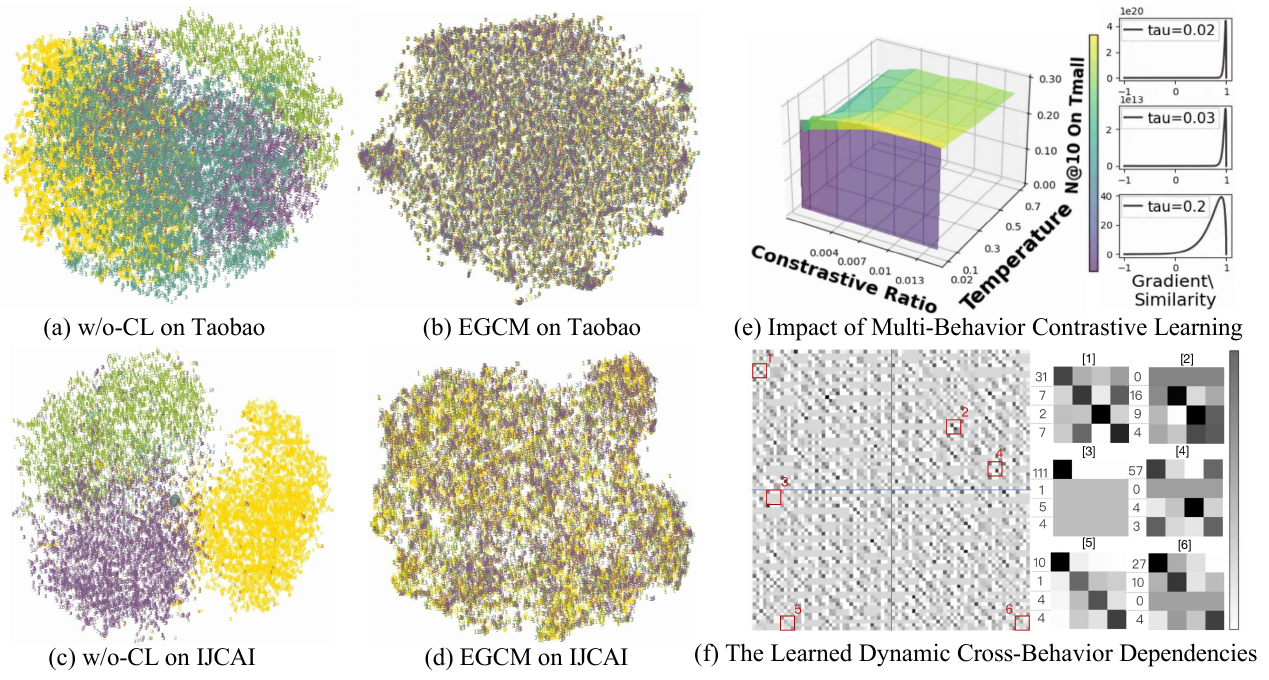}
    \vspace{-0.1in}
    \caption{(a)-(b): Visualized behavior-aware user embeddings which preserve multi-behavior commonality and cross-user diversity. (e): Impact study of contrastive learning in terms of temperature $\tau$. (f): Learned cross-behavior dependency weights across time slots.}
    \vspace{-0.1in}
    \label{fig:case_study}
\end{figure}

\vspace{-0.05in}
Our \model\ model is further evaluated from three perspectives. i) In Figure~\ref{fig:case_study} (a)-(d), we visualize the encoded behavior-specific user embeddings of our \model\ and compare them with the variant w/o-CL (without the multi-relational contrastive learning). The results demonstrate that our cross-behavior data augmentation scheme is effective in preserving users' multi-behavior commonality while better differentiating diverse user preferences. ii) We observe in Figure~\ref{fig:case_study} (e) that a smaller value of temperature value $\tau$ can lead to larger gradients, which can help identify hard negatives and enhance the model's discrimination ability in learning personalized user interests.
However, too small a temperature coefficient can result in gradient explosion, as shown in the purple part. 
iii) In Figure~\ref{fig:case_study} (f), we visualize the learned cross-behavior dependencies between time slots using $4\times 4$ weight matrices $\Phi$ in Eq.\ref{eq:self-attention-weights} for 400 sampled users on the Taobao dataset, and highlight six of them on the right side in Figure\ref{fig:case_study} (f). We observe that most matrices have the darkest diagonal color, which is the characteristic of self-attention. The right part shows that the weight is related to the number of interactions of the behavior. For example, user 22186-[3] the \emph{super node} has 111 interactions in the \emph{page view} behavior. However, the interactions in other behaviors are relatively too few (\ie, \{1, 5, 4\}) to learn differentiated values.

%% file: relate.tex
\vspace{-0.05in}
\section{Related Work}
\label{sec:relate}
\vspace{-0.05in}
\noindent \textbf{Sequential Recommendation}. 
Recent years have witnessed numerous efforts to leverage neural networks for modeling behavior dynamics in recommendation. RNN-based methods like GRU4Rec~\cite{hidasi2015session} and CNN-based approaches like Caser~\cite{tang2018personalized} have been proposed to capture the complex temporal dynamics in user-item interactions. Moreover, several self-attention models, such as SASRec~\cite{kang2018self}, BERT4Rec~\cite{sun2019bert4rec}, and TiSASRec~\cite{li2020time}, have been developed to estimate item correlations and capture long-range dependencies. Inspired by the power of GNNs, recent recommendation systems have applied graph-based message passing schemes to encode multi-order dependencies. Examples include MA-GNN~\cite{ma2020memory}, SURGE~\cite{chang2021sequential}, and GCE-GNN~\cite{wang2020global}. Furthermore, self-supervised learning has been used in recent sequential recommendation methods to augment the data, such as COTREC~\cite{xia2021self}, DHCN~\cite{xia2021self}. However, most existing methods focus on modeling a single type of interaction and ignore the heterogeneous user preferences.


\noindent \textbf{Multi-Behavior Recommender System}.
Multi-behavior recommendation has been demonstrated to greatly improve recommendation performance~\cite{chen2021graph}. For instance, NMTR~\cite{gao2019neural} and DPT~\cite{zhang2023denoising} differentiate behavior semantics using multi-task learning schemes. To capture the diverse relationships between users and items, recent studies such as MBGCN~\cite{jin2020multi}, MBGMN~\cite{xia2021graph}, and KMCLR~\cite{xuan2023knowledge} have leveraged GNNs to encode multi-behavior patterns.


\noindent \textbf{Contrastive Representation Learning}.
Contrastive learning has emerged as a popular technique for data augmentation with auxiliary self-supervised signals~\cite{ren2023sslrec,chen2023heterogeneous,liu2021self,wei2023multi}. In the domain of image analysis, numerous contrastive learning methods have been proposed for modeling image data with different augmentation techniques~\cite{aberdam2021sequence,verma2021towards,tian2020makes}, such as cropping, horizontal translations, and rotations. Cross-view contrastive learning has also been developed to achieve state-of-the-art graph representation performance using various augmentation operators, such as node shuffles in DGI~\cite{velickovic2019deep}, centrality-aware edge dropout in GCA~\cite{zhu2021graph}, and subgraph sampling in MVGLR~\cite{hassani2020contrastive}.

%% file: conclusion.tex
\vspace{-0.05in}
\section{Conclusion}
\label{sec:conclusion}
\vspace{-0.05in}
This work proposes a model called \model\ that captures the heterogeneity of each user's interactions at both the short-term and long-term interest levels. To better model the multi-behavior preferences of individual users and improve the distinction of behavior-aware preferences of different users, we introduce a multi-relational contrastive learning paradigm. Our experiments on three real-world datasets demonstrate that \model\ significantly outperforms various baselines in terms of recommendation performance. In the future, we plan to extend \model\ to adapt to cross-domain recommendation and tackle the cold-start problem by exploring LLMs for knowledge augmentation. This will enable us to leverage external knowledge sources to enhance the model's ability to make accurate recommendations.


%% file: appendix.tex
\section{Appendix}
\label{sec:Supplementary}

\subsection{Detailed Discussion on User Interest Representation Discrimination}
\label{sec:user_rep}
Contrastive loss is a hardness-aware loss function \cite{wang2021understanding, wu2021self} that can effectively push hard negative samples away from the anchor by giving them greater gradients under the contrastive training framework. This property is particularly beneficial for our multi-behavior graph neural architecture. One of the key challenges in existing GNN architectures is how to strike a balance between high-order connectivity modeling and the over-smoothing issue~\cite{li2018deeper}. Stacking more graph-based information propagation layers can increase the risk of over-smoothing and undermine the ability of the model to encode collaborative effects. Therefore, enhancing the discrimination ability of the user interest representation paradigm is necessary for recommender systems. To address this challenge, our multi-behavior contrastive learning framework assigns larger gradients to hard negative samples, thereby enhancing the discrimination of user representations.


\textbf{Embedding Normalization.}
To map embeddings with arbitrary value distributions into the same hyperspace, we perform embedding normalization as $\textbf{z}_i = \textbf{e}_i/\parallel \textbf{e}_i \parallel$, where $\textbf{e}_i$ denotes the output prior to normalization. The gradient of the loss with respect to $\textbf{e}_i$ is related to that with respect to $\textbf{z}_i$ via the chain rule, which is presented as follows:
\vspace{0.05in}
\begin{equation}
    \frac{ \partial{\mathcal{L}_i(\textbf{z}_i)} }{ \partial{ \textbf{e}_i }} =  \frac{ \partial{\mathcal{L}_i(\textbf{z}_i)} }{ \partial{\textbf{z}_i}}
    \frac{ \partial{\textbf{z}_i} }{ \partial{\textbf{e}_i} } 
\end{equation}
\begin{equation}
    \begin{split}
     \frac{\partial{\textbf{z}_i}}{\partial{\textbf{e}_i}} & = \frac{\partial{}}{\partial{\textbf{e}_i}}\left( \frac{\textbf{e}_i}{\parallel \textbf{e}_i \parallel} \right) 
                 = \frac{1}{\parallel \textbf{e}_i \parallel}I - \textbf{e}_i\left(\frac{\partial{(1/\parallel \textbf{e}_i\parallel)}}{\partial{\textbf{e}_i}} \right)^T\\
                 &= \frac{1}{\parallel \textbf{e}_i \parallel}\left(I - \frac{\textbf{e}_i \textbf{e}_i^T}{\parallel \textbf{e}_i \parallel^2}   \right) 
                 = \frac{1}{\parallel \textbf{e}_i \parallel}\left(I - \textbf{z}_i \textbf{z}_i^T \right)
    \end{split}
\end{equation}

\vspace{0.05in}
\textbf{Gradients of Negative Pairs.}
We use the loss function $\mathcal{L}_i$ to calculate the partial derivative of the anchor point and analyze the influence of different samples on the gradients:\vspace{0.05in}
\begin{equation}
    \label{eq:pos-neg1}
    \begin{split}
     \frac{\partial{L_i}}{\partial{\textbf{z}_i}} & =
     \frac{\partial{}}{\partial{\textbf{z}_i}} \left( 
     -log \frac{\exp{(\textbf{z}_i \cdot \textbf{z}_P / \tau)}}{ \sum_{\mathcal{V}_A}{\exp{(\textbf{z}_i \cdot \textbf{z}_A / \tau)}}}
     \right)
      = -\frac{1}{\tau} \cdot \textbf{z}_P + \frac{1}{\tau} \frac{\sum_{\mathcal{V}_A}{\textbf{z}_A \cdot \exp{(\textbf{z}_i \cdot \textbf{z}_A / \tau)}}}{\sum_{\mathcal{V}_A}{\exp{(\textbf{z}_i \cdot \textbf{z}_A / \tau)}}}\\
     & = \frac{1}{\tau} \left( 
     \frac{\sum_{\mathcal{V}_N}{\textbf{z}_N \cdot \exp{(\textbf{z}_i \cdot \textbf{z}_N / \tau)} }}{\sum_{\mathcal{V}_A}{\exp{(\textbf{z}_i \cdot \textbf{z}_A / \tau)}}}
     +\frac{\textbf{z}_P \cdot \left({ \exp{(\textbf{z}_i \cdot \textbf{z}_P / \tau)}}
     - \sum_{\mathcal{V}_A}{\exp{(\textbf{z}_i \cdot \textbf{z}_A / \tau)}} \right)
     }{\sum_{\mathcal{V}_A}{\exp{(\textbf{z}_i \cdot \textbf{z}_A / \tau)}} }
     \right) \\
    \end{split}
\end{equation}

\vspace{0.05in}
\noindent The two terms in the formula represent the gradient contributions of positive and negative samples, respectively. Here, $\textbf{z}_N$ represents an instance from the set of negative samples, while $\textbf{z}_P$ and $\textbf{z}_A$ are from the positive sample set and the entire set, respectively. We will focus on the negative part, which can be expressed as:
\vspace{0.08in}
\begin{equation}
    \label{eq:pos-neg2}
    \begin{split}
    & \frac{1}{\tau} \cdot \frac{\sum_{\mathcal{V}_N}{\textbf{z}_N \cdot \exp{(\textbf{z}_i \cdot \textbf{z}_N / \tau)} }}{\sum_{\mathcal{V}_A}{\exp{(\textbf{z}_i \cdot \textbf{z}_A / \tau)}}} 
     =  \frac{1}{\tau \cdot \parallel \textbf{e}_i \parallel} \cdot
    \frac{\sum_{\mathcal{V}_N}{ \left(\textbf{z}_N-(\textbf{z}_i \cdot \textbf{z}_N) \cdot \textbf{z}_i \right) \cdot \exp{(\textbf{z}_i \cdot \textbf{z}_N / \tau)} }}{\sum_{\mathcal{V}_A}{\exp{(\textbf{z}_i \cdot \textbf{z}_A / \tau)}}}    
    \end{split}
\end{equation}

\begin{figure}[H]
    \centering
    \includegraphics[width=0.68\columnwidth]{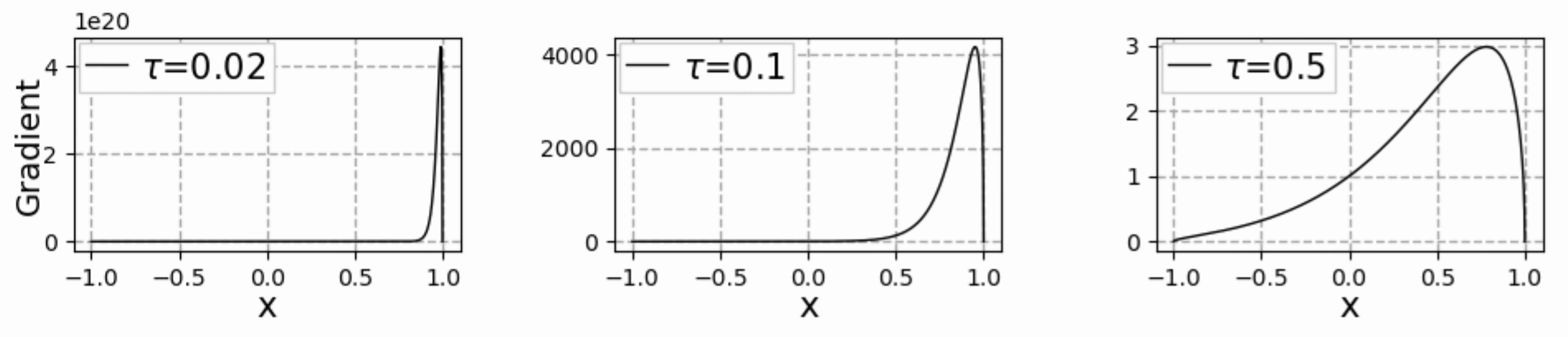}
    \vspace{-0.1in}
    \caption{ 
    Gradient function $c(x)$ in Eq.~\ref{eq:prop-relat1} when $\tau$ = 0.02, $\tau$ = 0.1 and $\tau$ = 0.5. $x$ is the similarity between positive and negative instances. This demonstrates that the gradient increases with decreasing temperature coefficient $\tau$. }
    \label{fig:gradient}
    \vspace{-0.1in}
\end{figure}

\vspace{0.05in}
\textbf{Temperature and Gradient.} The proportional term of the norm of the gradient of each term in the sum formula is:
\label{sec:supp-cl}
\begin{equation}
        \label{eq:neg}
    \begin{split}
\parallel & \textbf{z}_N-(\textbf{z}_i \cdot \textbf{z}_N)\cdot \textbf{z}_i  \parallel  \left| \frac{  \exp{(\textbf{z}_i \cdot \textbf{z}_N / \tau)} }{\sum_{\mathcal{V}_A}{\exp{(\textbf{z}_i \cdot \textbf{z}_A / \tau)}}} \right| \\
 &\propto \sqrt{1-(\textbf{z}_i \cdot \textbf{z}_N)^2} \cdot \exp{(\textbf{z}_i \cdot \textbf{z}_N / \tau)}
    \end{split}
\end{equation}
\noindent Since both $\textbf{z}_i$ and $\textbf{z}_N$ are unit vectors, we introduce another variable $x$ with the definition $x = \textbf{z}_i \cdot \textbf{z}_N \in [-1,1]$ to simplify the expression in Eq.~\ref{eq:neg}:
    \begin{equation}
        \label{eq:prop-relat1}
    \begin{split}
        c(x) \propto \sqrt{1-(x)^2} \cdot \exp{(x / \tau)}
    \end{split}
\end{equation}
\noindent $c(x)$ is the relationship function that determines the gradient contribution of negative samples. We illustrate the function in Eq.~\ref{eq:prop-relat1} in Figure~\ref{fig:gradient}, where the independent variable is the similarity $x$ and the dependent variable is proportional to the negative sample gradient. As $x$ increases, the gradient of negative samples also increases. Decreasing the temperature coefficient $\tau$ leads to a significant increase in the gradient of negative samples obtained via contrastive learning. 
In our recommended scenario, hard negative samples $\textbf{z}{N-hard}$ represent users other than the anchor user. If $\textbf{z}{N-hard}$ is very close to the anchor point $\textbf{z}_i$, the value of $x$ for the hard negative samples approaches 1, which results in more indistinguishable user representations. The contrastive loss gives a larger gradient, and as relative negative samples, the pairs will be pushed farther away from each other. In this way, \model\ enhances the user representations with multi-behavior diversity. Therefore, the experiments in Sec.~\ref{sec:in_depth_exp} show that as the temperature $\tau$ decreases, the user representations become more distinguishable and yield better performance. However, excessively small values of $\tau$ can cause gradient explosion. Therefore, the temperature should be within a proper range to achieve the best performance.

\subsection{ Dimensional Transformation of the Memory Module}
\begin{table}[H]
\centering
\caption{ Dimensional Details of Eq.~\ref{eq:self-attention-weights} for Self-attention of Multi-behavior Relationships.}
\begin{tabular}{l|l} \hline
\multicolumn{2}{c}{Dimensional Transformation of the Memory Module}  \\ \hline \hline
Parameters & Dimensionality \\ \hline \hline
Input         & ($|\mathcal{B}|$ $\times$ N $\times$ d) \\ \hline
Q,K,V Transformation     & ($|\mathcal{B}|$ $\times$ N $\times$ d) $\cdot$ (d $\times$ d) $\longrightarrow$  ($|\mathcal{B}|$ $\times$ N $\times$ d)   \\
Q Extension            & ($|\mathcal{B}|$ $\times$ N $\times$ d) $\longrightarrow$  ($|\mathcal{B}|$ $\times$ 1 $\times$ N $\times$ d)    \\ 
K Extension          & ($|\mathcal{B}|$ $\times$ N $\times$ d) $\longrightarrow$ (1 $\times$ $|\mathcal{B}|$ $\times$ N $\times$ d)         \\ 
V Extension          & ($|\mathcal{B}|$ $\times$ N $\times$ d) $\longrightarrow$ (1 $\times$ $|\mathcal{B}|$ $\times$ N $\times$ d)         \\ \hline
Self-attention          & ($|\mathcal{B}|$ $\times$ 1 $\times$ N $\times$ d) $\cdot$ (1 $\times$ $|\mathcal{B}|$ $\times$ N $\times$ d)  $\longrightarrow$   ($|\mathcal{B}|$ $\times$ $|\mathcal{B}|$ $\times$ N $\times$ d)     \\
Reduce Sum          & ($|\mathcal{B}|$ $\times$ $|\mathcal{B}|$ $\times$ N $\times$ d)       $\longrightarrow$   ($|\mathcal{B}|$ $\times$ $|\mathcal{B}|$ $\times$ N $\times$ 1)    \\
Softmax          & ($|\mathcal{B}|$ $\times$ $|\mathcal{B}|$(softmax) $\times$ N $\times$ 1)         \\\hline
Attention Matrix*V          & ($|\mathcal{B}|$ $\times$ $|\mathcal{B}|$ $\times$ N $\times$ d) $\cdot$ (1 $\times$ $|\mathcal{B}|$ $\times$ N $\times$ d) $\longrightarrow$ ($|\mathcal{B}|$ $\times$ $|\mathcal{B}|$ $\times$ N $\times$ d)    \\ 
Output          & ($|\mathcal{B}|$ $\times$ $|\mathcal{B}|$(reduce sum+squeeze) $\times$ N $\times$ d) $\longrightarrow$ ($|\mathcal{B}|$ $\times$ N $\times$ d)         \\\hline
\end{tabular}
\begin{tablenotes}
\footnotesize
\item{*} 'N' denotes the number of users or items.
\end{tablenotes}
\end{table}